\documentstyle[epsf]{l-aa}

\begin{document}

\title{Coronal X--Ray Emission of HD\,35850: the ASCA view}

\author{G.~Tagliaferri\inst{1} \and S.~Covino\inst{1}
\and T.A.~Fleming\inst{2}
\and M.~Gagn\'e\inst{3}
\and R.~Pallavicini\inst{4}
\and F.~Haardt\inst{5} \and Y.~Uchida\inst{6}}

\institute{Osservatorio Astronomico di Brera, via E. Bianchi 46, I--22055
Merate (LC), Italy
\and
Steward Observatory, University of Arizona, Tucson, Arizona
\and
JILA, University of Colorado, Boulder, CO 80309-0440, USA
\and
Osservatorio Astronomico G.S. Vaiana, Palermo, Italy
\and
Institute of Theoretical Physics, Chalmers University Technology,
412 96, G\"oteborg, Sweden
\and
Science University of Tokyo, Tokyo, Japan}

\thesaurus{08(08.01.1, 08.01.2, 08.03.5, 08.09.2: HD\,35850, 08.12.1,
13.25.5)}
\offprints{G.~Tagliaferri}

\date{Received: 06-09-1996; accepted: 21-10-1996}

\maketitle
                          
\label{sampout}

\begin{abstract}
We present the analysis of the X--ray data of the young active star
HD\,35850 obtained with ASCA and ROSAT. Our main goal was to see if
there is a difference in the elemental abundances of active stars
between young and more evolved objects.
A two temperature plasma with subsolar abundances, of the order of
$Z = 0.15 - 0.3 $, is required to fit the SIS spectra.
Similar results are obtained from a ROSAT PSPC observation.
Metal abundances of 0.2 -- 0.4 the solar value are required to
fit both the ASCA and ROSAT data together. From a simultaneous SIS0+SIS1
spectral fit, with 2T plasma models and
abundances free to vary in non-solar proportions, we find that, besides
N, O and Ne for which we find very low values, all other elements have
values relative to solar abundances 
around 0.2--0.3. These subsolar abundances are in line with those typically
observed in more evolved, active stars like RS CVn and Algol--type binaries.

The two temperature values required to fit the ASCA SIS spectra are
about 0.5 and 1.0 keV. These temperatures, especially  
the higher one, are lower with respect to the values found for the RS CVn
and Algol binaries or for the young star AB Dor, but higher than 
other single G/K stars. All our data show that this single, late F--type
star is actually a very active source, indirectly confirming that this
fast rotating star is probably a young object. In the simultaneous fit of
the ASCA+ROSAT data, a third temperature is required. However this is not
just an addition of a softer component, but is more a redistribution of the
dominant temperatures. Indeed, the range spanned by the three temperatures,
from 5 to 15 million degrees, is not very large.


\end{abstract}
 
\section{Introduction}

The decline of coronal emission with age is commonly
interpreted to be a consequence of reduced magnetic activity during
the course of stellar evolution. Mass loss through magnetized stellar
winds progressively decreases the star rotation rate, and this in turn 
makes the generation of surface magnetic fields by the dynamo process
less efficient. By studying X--rays from stars of different
age one has a powerful tool to investigate coronal emission under
different conditions of magnetic activity.

The spectral capabilities of ASCA now have the potential to study
in some detail the X--ray spectra of a sizable number of stars at
different levels of
activity and/or age. However, late--type stars in galactic clusters
and star--forming regions are typically too far away to be studied at
sufficiently high signal--to--noise ratio (S/N) with ASCA. Fortunately, a 
number of nearby isolated young stars which can be used as good proxies for  
cluster members, have been discovered recently. These objects
have been identified on the basis of various criteria, including
rapid rotation, enhanced X--ray emission (e.g. Vilhu et al. 1987),
large Li abundances (Pallavicini et al. 1992a, 1992b, Randich et al. 1993), 
space motions indicative of membership in associations (Innis et al. 1968, 
Anders et al. 1991), or a combination of the above criteria. The most
effective way to discover isolated young stars (i.e. NOT in star forming
regions) is probably through
X--ray and EUV surveys as demonstrated by optical
follow--up observations of serendipitous X--ray sources detected with
{\it Einstein}, EXOSAT, ROSAT and EUVE (Fleming et al. 1988, 1989;
Tagliaferri et al. 1992a, 1992b, 1994; Favata et al. 1993; Jeffries 1995).

We have carried out extensive photometric and spectroscopic observations of
the optical counterparts of stellar X--ray sources discovered serendipitously
by EXOSAT (Tagliaferri et al. 1992a, 1992b, 1994; Cutispoto et al. 1995) and
found that 
at least one 
third of the EXOSAT serendipitous sources are, in fact, young
stars with ages comparable to or younger than the Pleiades
($\sim 7 \times 10^7$ yrs).
Here we present an ASCA observation of one of these stars, HD\,35850.

HD\,35850 is a single star of spectral type F8/9~V 
with a spectroscopic distance $\sim 24$ pc (Cutispoto et al. 1995). The source 
was detected
serendipitously by EXOSAT with an intrinsic X--ray luminosity of
$\sim 1.6 \times 10^{30}$ erg s$^{-1}$. This value is quite remarkable
for a single, late F--type star and one might suspect the presence
of a less massive companion in order to explain its high X--ray
luminosity. However, our optical data seem to confirm that HD\,35850
is a true single star. We observed this source for six consecutive
nights taking high resolution spectra (R=50000, S/N$>100$) every night
in the H$_{\alpha}$ region. We found a constant radial velocity
(V$_{\rm r} = 24 \pm 2.4$ Km s$^{-1}$) over the six nights. This
leads us to exclude that the star is a short period spectroscopic
binary (Pastori et al. 1996).
The source has been detected again by both the ROSAT--WFC and EUVE
all--sky surveys, and as a serendipitous source in a ROSAT--PSPC
pointing (Panarella et al. 1996). The large Li abundance [$\log n(Li) =
3.2$, comparable to the protostellar Li abundance of Pop I stars],
solar metallicity, and large rotation rate ($v\sin i = 50$ km s$^{-1}$)
are all consistent with HD\,35850 being a young object, at least as young
as the Pleiades (Tagliaferri et al. 1994). 

\section{Observations and results}

The Advanced Satellite for Cosmology and Astrophysics (ASCA)
is an X--ray observatory carrying four detectors onboard, namely
two Solid State Imaging Spectrometers (SIS) and two Gas Imaging
Spectrometers (GIS). Each detector is at the focus of an imaging
thin foil grazing incidence telescope (Tanaka et al. 1994).
Each SIS has four CCD chips, but for our observation
they were operated in 1--CCD mode, that implies a field of view (FOV)
of $11^\prime \times 11^\prime$. The FWHM energy resolution of each SIS is
$\sim 60$--120 eV from 1--6 keV, compared to 200--600 eV for
the GIS. Each GIS has a $40^\prime$ diameter circular field of view.
SIS0 and GIS2 are the two best calibrated detectors.
The observation of HD\,35850 started on 1995 March 12 at
02:20 and ended at 19:20.

\subsection{Light Curves}

Using an extraction radius of 4 arcmin, the background subtracted source
count rates are $0.701 \pm 0.007$ for SIS0 and $0.200 \pm 0.004 \ cts\ s^{-1}$
for GIS2. The total good exposure time was 18 ksec.
The light curves from the 4 different detectors have been analysed.
In Fig.~1 we show the rebinned SIS light curves with a 1000 sec/bin stepsize.
Bins containing less than 500 sec of exposure time have been excluded.
In all four detectors, the count rate is not consistent with a constant
source, with flux variations of $\simeq 20$\% on timescales of a few hours
at the 3--$\sigma$ level in the two SIS datasets (Fig.~1). In order to test 
for a possible
energy dependence of the count rate
variation, we have produced separate light curves of the SIS datasets
for the low--energy channels (0.5--0.9 keV) and for the high--energy channels
(1.1--4 keV).  No significant change in the hardness ratio compatible
with the statistics of our data has been found. The statistical significance
of any flux change in the GIS datasets is lower, closer to 1--$\sigma$.

\begin{figure}[h]
\epsfxsize=8.8cm
\epsffile{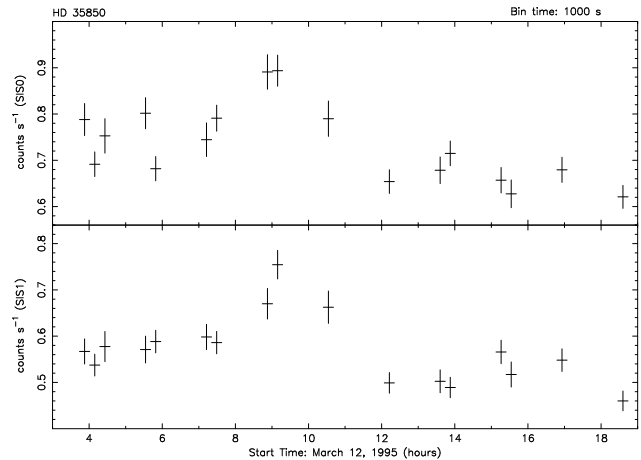}
\caption [] {SIS light curves of HD\,35850; a variability of the order
of $\sim 20\%$ is detected.}
\end{figure}

\subsection{Spectral Analysis}

Since the hardness ratio is essentially constant for the duration of the
observation, we have performed the spectral analysis using the whole data
set without any time filtering. 
The extracted spectra have been rebinned to give at least 25 counts per bin.
A background spectrum was accumulated from the outer regions of the field of 
view (FOV) for the GIS detectors. For the SIS detectors, we used the standard
background data provided by the ASCA observatory team since our source almost
completely fills the 1CCD mode FOV. Of course, this procedure can
only be followed if the background is constant, which
indeed was the case.

\begin{table*}
\begin{center}
\begin{tabular}{l|cc|cc|cc|c|cc}
\hline
{\bf SIS0} & $T_1$ & E.M.$_1$ & $T_2$ & E.M.$_2$ & $T_3$ & E.M.$_3$ & $Z$ & 
$\chi^2_{\nu}$ & {\it d.o.f.}\\
           & KeV   &$(10^{52} \, cm^{-3})$& KeV  &$(10^{52} \, cm^{-3})$
           & KeV  &$(10^{52} \, cm^{-3})$ \\ 
\hline
\\
RS  & $0.76 \pm ^{0.02}_{0.02}$ & $2.62$ & 
$1.75 \pm ^{0.33}_{0.23}$ & $1.71$ & $-$ & $-$ & 
$1$ & $2.89$ & 101\\
\\
 "  & $0.35 \pm ^{0.03}_{0.04}$ & $0.93$ & 
$0.80 \pm ^{0.03}_{0.03}$ & $2.11$ & 
$1.91 \pm ^{0.58}_{0.32}$ & $1.72$ & $1$ & $2.28$ & 99\\
\\
 "  & $0.44 \pm ^{0.19}_{0.10}$ & $2.87$ & 
$0.79 \pm ^{0.04}_{0.03}$ & $12.6$ & $-$ & $-$ & 
$0.15 \pm ^{0.03}_{0.03}$ & $1.37$ & 100\\
\\
MK  & $0.63 \pm ^{0.02}_{0.01}$ & $2.75$ & 
$1.94 \pm ^{0.31}_{0.23}$ & $1.94$ & $-$ & $-$ & 
$1$ & $2.60$ & 100\\
\\
 "  & $0.48 \pm ^{0.07}_{0.07}$ & $1.48$ & 
$0.77 \pm ^{0.08}_{0.06}$ & $1.62$ & 
$2.25 \pm ^{0.16}_{0.33}$ & $1.75$ & $1$ & $2.14$ & 99\\
\\
 "  & $0.63 \pm ^{0.02}_{0.08}$ & $11.7$ & 
$1.16 \pm ^{1.04}_{0.33}$ & $2.99$ & $-$ & $-$ & 
$0.18 \pm^{0.07}_{0.04}$ & $1.53$ & 100\\
\\
\hline
{\bf SIS1} & $T_1$ & E.M.$_1$ & $T_2$ & E.M.$_2$ & $T_3$ & E.M.$_3$ & $Z$ & 
$\chi^2_{\nu}$ & {\it d.o.f.}\\
           & KeV   &$(10^{52} \, cm^{-3})$& KeV  &$(10^{52} \, cm^{-3})$
           & KeV  &$(10^{52} \, cm^{-3})$ \\ 
\hline
\\
RS  & $0.51 \pm ^{0.16}_{0.04}$ & $1.04$ & 
$1.03 \pm ^{0.02}_{0.03}$ & $3.16$ & $-$ & $-$ & 
$1$ & $2.57$ & 92\\
\\
 "  & $0.43 \pm ^{0.08}_{0.08}$ & $0.72$ & 
$0.86 \pm ^{1.40}_{0.00}$ & $2.31$ & 
$1.75 \pm ^{0.39}_{0.40}$ & $1.44$ & $1$ & $2.19$ & 90\\
\\
 "  & $0.59 \pm ^{0.09}_{0.17}$ & $3.55$ & 
$0.91 \pm ^{0.00}_{0.05}$ & $8.22$ & $-$ & $-$ & 
$0.23 \pm ^{0.09}_{0.06}$ & $1.63$ & 91\\
\\
MK  & $0.64 \pm ^{0.02}_{0.02}$ & $2.36$ & 
$1.43 \pm ^{0.13}_{0.10}$ & $2.09$ & 
$-$ & $-$ & $1$ & $1.70$ & 92\\
\\
 "  & $0.59 \pm ^{0.05}_{0.09}$ & $1.93$ & 
$0.97 \pm^{1.33}_{0.24}$ & $1.09$ & 
$1.88 \pm ^{2.98}_{0.40}$ & $1.56$ & $1$ & $1.57$ & 90\\
\\
 "  & $0.63 \pm ^{0.02}_{0.04}$ & $6.04$ & 
$1.16 \pm ^{0.14}_{0.12}$ & $3.91$ & $-$ & $-$ & 
$0.32 \pm^{0.14}_{0.08}$ & $1.20$ & 91\\
\\
\hline
\end{tabular}
\end{center}
\caption[]{Fit parameters for SIS0 and SIS1. They are computed for RS and MK 
models, 2 or 3 components and with abundances fixed at the solar value
or free to vary in solar proportion. Errors are at $90$\% confidence
for 2 or 3 parameters of interest.}
\label{Tab:sis01}
\end{table*}
\begin{table*}
\begin{center}
\begin{tabular}{l|cc|cc|c|cc}
\hline
{\bf GIS2} & $T_1$ & E.M.$_1$ & $T_2$ & E.M.$_2$ & $Z$ & $\chi^2_{\nu}$
& {\it d.o.f.}\\
& (KeV) &($10^{52} \, cm^{-3}$)& (KeV)& ($10^{52} \, cm^{-3}$)& \\
\hline
\\
RS  & $0.65 \pm ^{0.12}_{0.16}$ & $1.62$ & $1.30 \pm 
^{0.34}_{0.22}$ & $1.94$ & $1$ & $0.79$ & 111\\
\\
 "  & $0.73 (\pm ^{0.07}_{0.07})$ & $13.8$ & $-$ & $-$ & $0.12 (\pm 
^{0.09}_{0.05})$ & $0.68$ & 112\\
\\
MK  & $0.61 \pm ^{0.08}_{0.09}$ & $2.20$ & $1.53 \pm 
^{0.32}_{0.20}$ & $1.73$ & $1$ & $0.81$ & 111\\
\\
"   & $0.68 \pm ^{0.07}_{0.07}$ & $16.4$ & $-$ & $-$ & $0.11 \pm 
^{0.06}_{0.04}$ & $1.05$ & 112\\
\\
\hline
{\bf GIS3} & $T_1$ & E.M.$_1$ & $T_2$ & E.M.$_2$ & $Z$ & $\chi^2_{\nu}$ \\
\hline
\\
RS & $0.78 \pm^{0.04}_{0.45}$ & $2.39$ & $1.69 \pm^{1.88}_{0.57}$ 
& $1.40$ & $1$ & $0.78$ & 125\\
\\
"  & $0.80 \pm^{0.04}_{0.05}$ & $10.7$ & $-$ & $-$ & $0.21 
\pm^{0.12}_{0.08}$ & $0.78$ & 126\\
\\ 
MK & $0.62 \pm^{0.09}_{0.11}$ & $2.38$ & $1.56 
\pm ^{0.40}_{0.23}$ & $1.95$ & $1$ & $0.74$ & 125\\
\\
"  & $0.74 \pm^{0.07}_{0.06}$ & $1.36$ & $-$ & $-$ & $0.16 
\pm^{0.08}_{0.05}$ & $0.83$ & 126\\
\\
\hline
\end{tabular}
\end{center}
\caption[]{Fit parameters for GIS2 and GIS3. They are computed for RS and
MK models, 1 or 2 components and with abundances fixed at the solar value
or free to vary in solar proportion. Errors with $90$\% confidence for
2 parameters of interest.}
\label{Tab:g23}
\end{table*}

While analyzing our ASCA data, we encountered several problems. 
In particular, although we tried various procedures, we never
obtained statistically fully satisfactory spectral fits
for the SIS detectors. This could be partly attributed to uncertainties
in some of the models used (see below) and/or in detector calibration.
We now give a summary of the tests we performed during the data
analysis. Then we will concentrate on the results obtained.
In order to estimate the uncertainties on
the PSF calibration, we extracted SIS spectra using 
two different circle sizes, i.e. with a 
smaller circle implying a larger PSF correction. We used a
radius of 38 original pixels and another of 150 original pixels (130 for SIS1).
The larger circle is the one suggested by the ASCA observatory team (for SIS1
we had to use a slightly smaller radius as the source is not centered in the 
1CCD detector FOV). All fits
with different models, detector combinations, and energy
ranges (by ignoring channels) were performed on both spectra extracted using
both circle sizes. As a result, we found that the best--fit values for the
temperatures and abundances are quite similar in the two cases, but they are
better constrained for the larger extraction radius,
although in this case the fits are statistically less
satisfactory (i.e. the reduced $\chi ^2$ is always larger).
The formally better fits obtained with the smaller extraction radius 
are probably due to the larger statistical errors associated with the
energy bins, given that using a smaller circle implies fewer counts per bin. 
The higher S/N
spectra amplify the uncertainties in the SIS calibration and in the spectral
models. Also, we should point out that two
different normalizations for every component are necessary to fit 
the two SIS spectra
extracted in the smaller circle (one for each detector), while
they are not required for the larger circle SIS spectra.
This immediately shows that the PSF uncertainties are reduced by using
the larger circle. In the following discussion, we will concentrate only
on the larger circle SIS spectra.

The low--energy channels contribute most to the $\chi ^2$.
It is already known that there are particular problems with the calibration
of the SIS detectors below 0.55 keV. To test the influence of the low--energy 
channels on the spectral fit results,
we systematically fitted the SIS0 spectrum by gradually increasing the 
low--energy
cut-off (SIS0 is the better calibrated between the two SISs).
The lowest energy values considered in the various cases are 0.3, 0.55, 0.7 and
0.8 keV. We found that the reduced $\chi ^2$
significantly improves each time we removed the lowest energy bins. The
elemental abundance value in solar proportion (Anders \& Grevesse 1989)
goes from essentially solar to very low value as 
the low--energy limit goes from 0.3 to 0.55 keV. Although any further 
increase of the 
low--energy limit does not significantly change the abundance value, it does 
cause a change in the temperature values.
In fact, the second temperature component is not constrained any longer 
and, indeed, a single intermediate
temperature is sufficient to fit the data. Based on these results, we decided
to use 0.55 keV as the lower energy limit.

A similar problem arose for the channels above 4 keV. In all four
detectors, the count rates in the high--energy channels are systematically in 
excess of  
the various best--fitted models. Being common to all four 
detectors, such 
an excess could be intrinsic to the source spectrum, but it could also
be due to uncertainties in the data and/or in the calibration.
As a matter of fact, there are few bins above 4 keV and they have large 
errors due to the low count rate at high--energy (we rebinned the
spectrum to have at least 25 counts per bin). 
We then decided to ignore all the energy channels above 4 keV, since   
they have a limited influence on the fits. 

Finally, we computed the errors in our model fitting using 
two different statistical methods. In our spectra, the counts
in each bin go from 25 to a few hundred and standard Poissonian statistics
should be fairly adequate. However, these counts also include the
background contribution. Thus, in the background subtracted spectra,
there could be some bins with too low a count rate to properly apply 
Poissonian statistics. Special statistics with $\sigma=(1+\sqrt{N+0.75})$
is found to be better suited  when the number of counts $N$ is small
(Gehrels 1986). When N is large, the two are essentially the same. We
performed extensive tests with the XSPEC package (version 9.01) using
statistical errors computed with the two procedures and we found that
the best--fit parameters in non--critical
situations (e.g. when the best--fit parameters are well defined)
are almost identical, although in some special cases they are 
better constrained using the second statistical method.
In particular, this non--standard way of weighting, as expected, is 
less sensitive to the channels above $\sim 4 $ keV, which have the
lowest number of counts. In addition, $\chi ^2$ is lower, but again
this is not surprising since the error bars are larger. 
In the following, we use the second statistical method. However, it is worth 
pointing out that no difference in the results obtained with the 
two statistical methods was found after we restricted the SIS intervals
between 0.55--4 keV.

For the spectral analysis, we used the optically thin plasma models by
Raymond \& Smith (RS, 1977) and Mewe et al. (MK, 1996a), also known as
``mekal", as implemented inside XSPEC. Emission measures were defined as 
E.M. = $\int n^2\,dV$ and E.M. = $\int n_en_H\,dV$ for RS and MK models,
respectively.
In all models, the column density $N_{H}$ was fixed at $1 \times 10^{18}$ 
cm$^{-2}$ as determined from the EUVE data (Gagn\'e et al. 1996). 
A two-temperature model with solar abundances gives unacceptable
fits to the SIS spectra. A clear improvement is obtained either by
adding a third temperature component or by allowing
the abundances to vary (in solar proportion). The latter
model gives much better fits with subsolar abundances 
($Z \simeq 0.15 - 0.3$; see Table\,\ref{Tab:sis01}),
while the former is still unacceptable. In figure 2, we plot the SIS0
spectrum together with the best--fitted 3T MK model with abundances
fixed at the solar value. Note the structures in the residuals, which
immediately show that a solar value for the abundances does not allow
a good fit to the data.
However, in no case are our fits to the data fully satisfactory (see
Table\,\ref{Tab:sis01}). Given the statistics of our data, we did not
try a 3T model with variable abundances.

The spectra obtained with the two GIS cameras have lower spectral resolution
and are also noisier.
A good fit to the GIS data is obtained both with $2T$ models with
solar abundances or with $1T$ models with lower--than--solar abundances. 
In the latter case, the
abundance values  are very similar to those obtained with the SISs
(Table \ref{Tab:g23}).

\begin{figure}
\epsfxsize=8.8cm
\epsffile{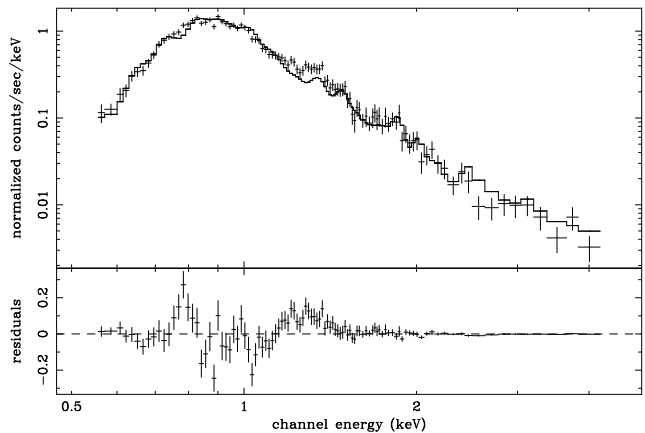}
\nopagebreak
\caption [] {Fit of the best calibrated SIS detector, SIS0, with a
3--temperature MK model and abundances fixed to the solar value. Note
the structures in the residuals; assuming solar abundance clearly
does not fit the data even with a 3T model.}
\end{figure}

We then performed simultaneous fits by combining the best
calibrated detectors, namely SIS0+GIS2, or the spectra with the highest 
resolution, namely SIS0+SIS1.
In the first case, we considered 2T and 3T models with metal abundances
either fixed to solar or free to vary in solar proportions. The
inclusion of more spectral bins seems to formally improve the average quality
of the fits as measured by the $\chi^2$ values. This analysis
confirms the results obtained with SIS0 alone. The fit parameters are
essentially the same with somewhat smaller 90\%--confidence limits. 
In the second case, SIS0+SIS1, we considered a $2T$ model with variable
abundances.
As can be seen from Table~\ref{Tab:sis0sis1}, there is quite good
agreement between the RS and MK models. They both have a very low value
of N, O and Ne. All other elements are around 0.2--0.3, except 
Ni, which has a higher value in the RS model while it is similar 
to the other elements in the MK model.

\begin{table}
\begin{center}
\begin{tabular}{l|cc}
\hline
\\
 & RS & MK \\
\hline
\\
 $T_1$\,(KeV)   & $0.50\pm^{0.03}_{0.02}$ & $0.57\pm_{0.04}^{0.02}$   \\
 E.M.$_1$\,($10^{52} \, cm^{-3}$) & $6.56$     & $7.19$                    \\
 $T_2$\,(KeV)   & $0.95\pm^{0.02}_{0.11}$ & $1.00\pm_{0.08}^{0.05}$   \\
 E.M.$_2$\,($10^{52} \, cm^{-3}$) & $7.26$           & $5.07$  \\
 N              & $0\,(<0.74)$            & $0\,(<2.33)$              \\
 O              & $0.04\pm^{0.00}_{0.04}$ & $0.06\pm_{0.05}^{0.05}$   \\
 Ne             & $0.06\pm^{0.00}_{0.06}$ & $0.01\pm_{0.01}^{0.06}$   \\
 Mg             & $0.19\pm^{0.07}_{0.07}$ & $0.35\pm_{0.08}^{0.06}$   \\
 Si             & $0.18\pm^{0.03}_{0.06}$ & $0.23\pm_{0.05}^{0.07}$   \\
 S              & $0.17\pm^{0.00}_{0.12}$ & $0.24\pm_{0.14}^{0.14}$   \\
 Fe             & $0.17\pm^{0.04}_{0.04}$ & $0.25\pm_{0.04}^{0.03}$   \\
 Ni             & $1.18\pm^{0.37}_{0.38}$ & $0.39\pm_{0.26}^{0.32}$   \\
 $\chi^2_{\nu}$ & 1.42                    & 1.33                      \\
 {\it d.o.f.}   & 189                     & 189                       \\
\hline
\end{tabular}
\end{center}
\caption[]{Fit parameters for SIS0+SIS1. They are computed for RS and MK 
models with 2 components and variable abundances. 
Abundances errors are based on $\chi^2_{min} + 2.71$.}
\label{Tab:sis0sis1}
\end{table}

\begin{table*}
\begin{center}
\begin{tabular}{l|cc|cc|c|cc}
\hline
{\bf ROSAT} & $T_1$ & E.M.$_1$ & $T_2$ & E.M.$_2$ & $Z$ & $\chi^2_{\nu}$ 
& {\it d.o.f.}\\
& (KeV) &($10^{52} \, cm^{-3}$)& (KeV)& ($10^{52} \, cm^{-3}$)& \\
\hline
\\
RS & $0.20 \pm ^{0.04}_{0.03}$ & $1.31$ & $0.80 \pm 
^{0.06}_{0.05}$ & $3.16$ & $1$ & $1.64$ & 48\\
\\
"  & $0.63 \pm ^{0.07}_{0.37}$ & $6.98$ & $1.28 \pm 
^{\infty}_{0.50}$ & $3.08$ & $0.29 \pm_{0.06}^{0.25}$ & $1.39$ & 47\\
\\
MK & $0.20 \pm ^{0.04}_{0.04}$ & $1.15$ & $0.73 \pm 
^{0.10}_{0.05}$ & $3.41$ & $1$ & $1.41$ & 48\\
\\
"  & $0.47 \pm ^{\infty}_{0.26}$ & $4.79$ & $0.93 \pm 
^{\infty}_{0.00}$ & $4.11$ & $0.42\pm^{0.23}_{0.10}$ & $1.23$ & 47\\
\\
\hline
\end{tabular}
\end{center}
\caption[]{Fit parameters for the ROSAT PSPC data. They are computed 
for RS and MK 
models with abundances fixed at the solar value or free to vary in
solar proportion. Errors with $90$\% confidence are given for 2 or 3 parameters
of interest.}
\label{Tab:rosat}
\end{table*}

\begin{table*}
\begin{center}
\begin{tabular}{l|ll|ll}
\hline
RS & Free & normaliz. & ASCA norm. ~~= & ROSAT norm. \\
\hline
   & Solar ab. & Non--sol. ab. & Solar ab. & Non--sol. ab. \\
\\
$T_1$            & $0.20\pm_{0.01}^{0.06}$ & $0.35\pm_{0.06}^{0.10}$ & 
$0.19\pm_{0.03}^{0.01}$ & $0.36\pm_{0.04}^{0.08}$ \\
E.M.$_1$\,(SIS0+GIS2)  & $0.75\times10^{52}$     & $2.21\times10^{52}$ 
    & $1.17\times10^{52}$     & $2.32\times10^{52}$     \\
E.M.$_1$\,(PSPC)& $1.25\times10^{52}$     & $1.94\times10^{52}$     & 
$-$		       &  $-$			 \\
$T_2$            & $0.74\pm_{0.02}^{0.03}$ & $0.68\pm_{0.04}^{0.07}$ &
 $0.74\pm_{0.04}^{0.03}$ & $0.79\pm_{0.06}^{0.04}$ \\
E.M.$_2$\,(SIS0+GIS2)  & $2.18\times10^{52}$     & $4.45\times10^{52}$
     & $2.23\times10^{52}$     & $5.87\times10^{52}$     \\
E.M.$_2$\,(PSPC)& $2.34\times10^{52}$     & $8.92\times10^{52}$     & 
$-$		       &  $-$			 \\
$T_3$            & $1.65\pm_{0.18}^{0.33}$ & $0.99\pm_{0.05}^{1.13}$ &
 $1.65\pm_{0.26}^{0.33}$ & $1.61\pm_{0.44}^{1.35}$ \\
E.M.$_3$\,(SIS0+GIS2)  & $1.56\times10^{52}$     & $4.69\times10^{52}$ 
    & $1.56\times10^{52}$     & $1.43\times10^{52}$     \\
E.M.$_3$\,(PSPC)& $1.60\times10^{52}$     & $1.20\times10^{52}$     & 
$-$		       &  $-$                    \\
$Z$              & $1$                     & $0.25\pm_{0.03}^{0.05}$ &
 $1$                     & $0.34\pm_{0.05}^{0.04}$ \\
$\chi^2_{\nu}$   & 2.25                    & 1.73		     &
 2.45                    & 1.84\\
{\it d.o.f.}     & 258                     & 257		     &
 261                     & 260\\
\\
\hline
MK & Free & normaliz. & ASCA norm. ~~= & ROSAT norm. \\
\hline
   & Solar ab. & Non--sol. ab. & Solar ab. & Non--sol. ab. \\
\hline
\\
$T_1$            & $0.17\pm_{0.04}^{0.02}$ & $0.53\pm_{0.09}^{0.07}$   
& $0.15\pm_{0.04}^{0.02}$ & $0.52\pm_{0.07}^{0.07}$   \\
E.M.$_1$\,(SIS0+GIS2)  & $0.33\times10^{52}$     & $4.37\times10^{52}$ 
      & $0.68\times10^{52}$     & $4.64\times10^{52}$       \\
E.M.$_1$\,(PSPC)& $0.72\times10^{52}$     & $4.76\times10^{52}$       &
 $-$			 & $-$			     \\
$T_2$            & $0.63\pm_{0.02}^{0.02}$ & $0.78\pm_{0.10}^{0.19}$   &
 $0.62\pm_{0.02}^{0.02}$ & $0.78\pm_{0.09}^{0.15}$   \\
E.M.$_2$\,(SIS0+GIS2)  & $2.63\times10^{52}$     & $3.30\times10^{52}$  
     & $2.75\times10^{52}$     & $3.64\times10^{52}$       \\
E.M.$_2$\,(PSPC)& $2.95\times10^{52}$     & $4.26\times10^{52}$       & 
$-$			 & $-$			     \\
$T_3$            & $1.78\pm_{0.16}^{0.23}$ & $1.83\pm_{0.36}^{0.85}$   &
 $1.80\pm_{0.16}^{0.23}$ & $1.90\pm_{0.38}^{1.00}$   \\
E.M.$_3$\,(SIS0+GIS2)  & $1.71\times10^{52}$     & $1.71\times10^{52}$  
     & $1.69\times10^{52}$     & $1.57\times10^{52}$       \\
E.M.$_3$\,(PSPC)& $1.60\times10^{-3}$     & $0.57\times10^{-3}$       & 
$-$			 & $-$			     \\
$Z$              & $1$                     & $0.35\pm_{0.04}^{0.06}$   &
 $1$                     & $0.34\pm_{0.03}^{0.04}$   \\
$\chi^2_{\nu}$   & 2.03                    & 1.60                      &
 2.18                    & 1.70                      \\
{\it d.o.f.}     & 258                     & 257                       &
 261                     & 260                       \\
\\
\hline
\end{tabular}
\end{center}
\caption[]{Fit parameters for SIS0+GIS2+PSPC. They are computed for RS and MK 
models, with 3 components and abundances fixed at the solar value or free to
vary in solar proportion. We considered the case with free normalizations
between the ROSAT and ASCA data for each component and the case with a single
normalization for both datasets for each component.
Errors with 90\% confidence are given for 3 or 4 parameters of
interest.}
\label{Tab:s0g2RO}
\end{table*}

\subsection{ROSAT data}

We also considered ROSAT PSPC data of HD\,35850 taken from the public
archive. HD\,35850 was serendipitously detected during a PSPC observation
of more than 5000\,s carried out in March 1992. It had a PSPC count rate
of $ 2.8 \ cts \ s^{-1}$ for a total of more than 14000 counts in the
spectrum. We fitted the PSPC spectrum with a 2T (RS and MK) model plus
the $N_H$ fixed at the EUVE value. We note that ROSAT, in contrast with ASCA,
is quite sensitive to the $N_H$, so the assumed $N_H$ value has an important
impact on the other parameters. We could have also chosen to fit this
parameter but, if we wish to allow the abundances to
vary, then there would be too many free parameters given the (low) statistics
of our spectrum and the poor ROSAT spectral resolution. Since EUVE is very
good in determining the $N_H$, we preferred to use the $N_H$ value as
determined with EUVE. As can be seen from Table~\ref{Tab:rosat}, if we
assume solar abundances, statistically, the best--fit is not fully satisfactory
with a reduced $\chi ^2$ of 1.6 and 1.4 with RS and MK respectively
(48 degree of freedom -- dof). If we allow the abundance to vary in
solar proportion, the best--fit value for the abundance is  0.3--0.4
for the two models, respectively. An F--test shows that the improvement of
the $\chi ^2$ is significant at more than 99\%. However, the two
temperatures, which increased, are not well constrained any more by the
ROSAT data alone. It is in any case encouraging that both the temperatures
and the abundance values are in good agreement with the ASCA ones.
However, with the abundances fixed to solar value, the cooler ROSAT
temperature is a factor of three lower than the ASCA one. Similar
temperature values, with assumed solar abundances, are commonly
found for active stars with ROSAT (e.g. Dempsey et al. 1993).

The 0.5--2.0 keV source flux in the two observations (ASCA and
ROSAT) is at similar levels within the errors,
i.e. $\sim 1.3$ and $1.4 \times 10^{-11}$ erg cm$^{-2}$
s$^{-1}$ respectively, corresponding to an X--ray luminosity of 
$\sim 1.5 \times 10^{30}$ erg s$^{-1}$ in the 0.5--2.0 keV energy band. 
Thus, we decided to simultaneously fit the PSPC+SIS0+GIS2 data of HD\,35850
(Fig.~3). To fit the data we had to use a 3T model, while the metal
abundances were either kept fixed at solar values or free to vary in solar
proportion. Because the ROSAT and ASCA observations are not simultaneous,
we also allowed different ROSAT and ASCA normalizations.
The results are reported in
Table~\ref{Tab:s0g2RO}. It is clear that a plasma with solar abundances
can not fit the data. Adding together the RS and MK intervals, at a 90\%
confidence level, the metal abundance value is determined to be between
0.2--0.4. The two cooler temperatures are well--constrained, while
the warmest component at 1.6--1.9\,keV is less precisely defined
(Table~\ref{Tab:s0g2RO}). If we allow for different normalizations
between ROSAT and ASCA, then an F--test shows that the best--fit improvement
is statistically significant at more than a 99\% level in all cases (RS and MK,
solar and non--solar abundances). However, all other parameters (temperatures
and abundances) are essentially the same (see Table~\ref{Tab:s0g2RO}).
Since the fluxes estimated during the two observations are similar and since
the ratios of the three normalizations between the two missions are different
with no systematic trend, the need of different normalizations between ROSAT
and ASCA is probably due to detector calibration uncertainties rather than to
source variability.
\begin{figure*}
\epsfxsize=18.0cm
\epsffile{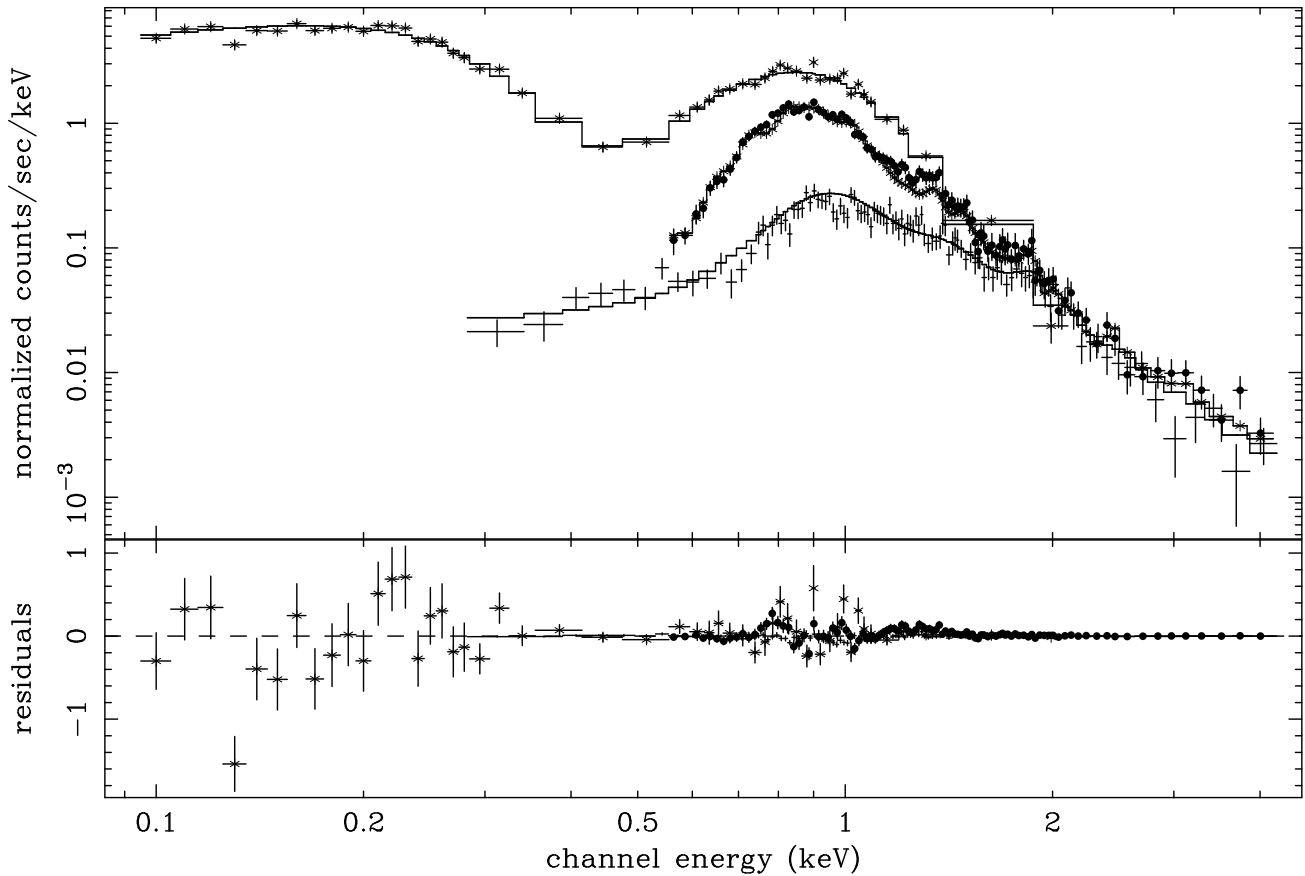}
\caption [] {Simultaneous fit of the two ASCA SIS0+GIS2 spectra and
non-simultaneous ROSAT PSPC data with a 3T MK model with abundances
free to vary in solar proportion. In this case the best--fit value is
$Z = 0.35$.}
\end{figure*}

\section{Discussion}

Our main goal was to see if there is a difference in the elemental 
abundances between very active young stars, as HD\,35850 should be,
and more evolved active stars such as the RS CVn--type binaries.

In the fits of the SIS spectra, all models with solar abundance
(2T and 3T RS or MK) that we considered do not give acceptable fits.
Instead, a 2T model with abundances free to vary in solar proportion
gives much better fits, although statistically not fully
satisfactory, with most of the $\chi ^2$ contribution arising from the
structures between 0.7 and 1.1 keV. This is probably due to either an 
inaccurate calibration of the SIS detectors or to uncertainties and/or
inaccuracies in the current plasma models, which are below
the noise in the low--energy resolution GIS spectra, but could well affect 
the higher resolution SIS data.
In fact, there is growing evidence that plasma codes fail to some degree to 
model the cool star spectra obtained with both the EUVE and ASCA
satellites. Theoretical and/or numerical improvements are thus required in
the commonly used thin plasma models (for a full discussion see Brickhouse
et al. 1995). In any case, subsolar abundances on the order of
$Z = 0.15 - 0.3 $ are required to fit our ASCA--SIS data. 

Not surprisingly, the analysis of the GIS data gives less clear results due to
the lower spectral resolution and S/N. Statistically, the GIS data
can equally be fitted
by a 1T plasma with subsolar abundances or by a 2T plasma with abundances
fixed at solar value. On the other hand, the ROSAT PSPC data seem to require
a 2T model with subsolar abundances, although from these data alone
it is not possible to properly constrain the temperatures, while the
abundances are loosely determined to be between 0.2--0.6 the solar value.
These results have been improved by simultaneous spectral fitting of data
from different detector combinations. Analyzing together the best calibrated
detectors, i.e. SIS0 and GIS2, simply confirms the results obtained
with the SIS0 fits. We then widened the energy range by simultaneously
fitting SIS0, GIS2 and ROSAT PSPC data (not simultaneous with the ASCA data).
In this case, a third temperature is required to fit the data, confirming
the results already found for other sources (e.g. Singh et al. 1995).
We used 3T models with fixed solar abundances or abundances variable in
solar proportion. Once again, we find that subsolar metal
abundances, with values between 0.2 and 0.4, 
are required to fit the ROSAT+ASCA data. We can then conclude that all of
our ASCA and ROSAT data show that the coronal plasma of HD\,35850 is
characterised by metal abundances lower than solar\footnote{We caution
that the preliminary results presented by us elsewhere (which indicated a
solar abundance for HD35850; Tagliaferri et al. 1996a,b) were based on a
simplified analysis which included the channels below 0.55 keV. As discussed
in section 2.2, this has a crucial effect on the derived abundances.},
on the order of $Z \approx 0.3$. This is in line with previous ASCA results.
As a matter of fact, the ASCA data of active cool stars analysed so far
point towards abundances typically in the range 0.1--0.3 of the solar value.
These ASCA results are valid, in general, for very active and more evolved
stars like the RS CVn and Algol binaries (White et al. 1994, White 1996;
Singh et al. 1995, 1996; Ortolani et al. 1996). Nevertheless, low abundances
have also been found in flare stars (Gotthelf et al. 1994, Mewe et al. 1996c)
and, to a lesser extent, in single
G/K--type stars (Drake et al. 1994). Furthermore, there is the case of AB Dor,
a young star whose photospheric abundances are solar (Vilhu et al. 1987), but
for which a metal poor corona, a factor 2--3 below solar, is required to fit
simultaneous ASCA and EUVE spectra (White et al. 1996, Mewe et al. 1996b).
And now we have the case of HD\,35850, a young star whose photospheric
abundances are also solar (Tagliaferri et al. 1994), but the corona is
subsolar. So the general trend seems to be that more evolved, very active
stars have low coronal metal abundances, while for the solar-type star
EK Dra probably this is not the case (G\"udel et al. 1996).
These conclusions are not based on ASCA data alone, but also on GINGA
(Tsuru et al. 1989; Stern et al. 1992) and, more recently, on ROSAT and
EUVE data (e.g. Ottmann \& Schmitt 1996; K\"urster \& Schmitt 1996; Singh
et al. 1996; Stern et al. 1995; Rucinski et al. 1995; Mewe et al. 1996c;
Schmitt et al. 1996).

The low spectral resolution of present X-ray detectors is not sufficient
to resolve individual spectral lines, particularly in the Fe-L complex
near 1 keV.  This, combined with incomplete and possibly incorrect model
line emissivities, can lead to erroneous abundance determinations (e.g.
Liedahl et al. 1995) even if the effect is not so large to explain a reduction
factor 3--5 in the abundances. However, all evidence points towards low metal
abundances in very active stars. This seems to be a robust result, as 
it can not be fully ascribed to detector calibration problems.
A tentative explanation can be given 
recalling the well known fact that coronal abundances in
very active regions of the Sun are quite different from those in 
the photosphere. In particular, there is the so-called FIP (First Ionization
Potential) effect. Elements having FIPs less than $\sim 10$ eV
are overabundant by typically a factor of four compared to those
elements with
higher FIPs (Meyer 1991; Feldman 1992). More recently, data obtained
with the spacecraft Ulysses have shown that the FIP effect is much reduced
in inactive regions (e.g. coronal holes) (Geiss et al. 1995).
So already in the Sun there is a clear link between activity and metal
abundances in the coronal plasma, but the situation is not at all clear and
differences between different coronal structures or over time in the same
regions have been detected (see Haisch et al. 1995 for a full discussion). 
However, in the case of active stars, no FIP effect has been seen so far
(but see G\"udel et al. 1996; Mg overabundant by a factor $\sim 2$,
while the other elements seem solar).
Rather a general underabundance of most elements seems to prevail. 
This leads us to ask if there is a ``saturation effect"
also with respect to coronal elemental abundances.

To check if a FIP effect is present in the SIS data of HD\,35850, we 
performed simultaneous SIS0+SIS1 spectral fits with 2T plasma models
and abundances free to vary in non-solar proportions.
We found quite a good agreement between the RS and MK models,
with very low values of N, O and Ne. All other elements are around
0.2--0.3, except Ni, which has a high value in the RS model
while it is more in line with the other elements in the MK model.
So there could be an indication that elements with high FIP (N, O, Ne)
are more depleted than the element with low FIP. However, the 
uncertainties both in the models and in the detector calibrations are
such that we can not draw any conclusion at the present stage. 
In any case we note that, besides activity, age should also play a role,
i.e. coronal elemental abundances are
probably related both to the age and to the activity level of the star.
For instance, the old but active Pop II binary HD\,89499 presents a
virtually metal free corona, i.e. still lower coronal abundances (Fleming
\& Tagliaferri 1996), in line with its very low photospheric metallicity
due to its age. We may conclude by saying that the picture of metal
abundance behaviour in the corona of active stars is still far from clear.
To improve our understanding, we probably have to wait for the next
generation of X-ray detectors with higher energy resolution and larger
effective area, such as the gratings that will be on board the AXAF and
XMM satellites.

As expected, two temperatures are necessary to fit the ASCA spectra and
they seem very well constrained. If we consider the RS and MK results 
together for the various detectors, the softer component is between T =
0.4 --  0.7 keV, while the harder one is between T = 0.8 -- 1.1 keV.
These results are confirmed by the analysis of the ROSAT data, although
with less constrained limits. These temperature values, in particular
the hotter ones, are lower with respect to the values found for the RS CVn
and Algol binaries (White et al. 1994; Antunes et al. 1994; Singh et al.
1995, 1996) and for AB Dor (White et al. 1996; Mewe et al. 1996b), but
higher than other single G/K stars (Drake et al. 1994). Not only the
plasma temperatures but also the total luminosity of this star is
remarkable. The 0.5--2.0 keV luminosity is above $10^{30} \ erg 
\ s^{-1}$  in all of the three X-ray detections (EXOSAT, ROSAT and
ASCA, spanning 9 years in time), an extremely high value
for a single late F-type star (e.g. Fleming et al. 1989, 1995;
G\"udel et al. 1995; Schmitt 1996). This is an indirect confirmation that
HD\,35850 is a young, rapidly rotating object.

In the simultaneous fit of the ASCA+ROSAT data, a third temperature
is required. However, this is not just an addition of a softer component,
but is more a redistribution of the dominant temperatures, with a widening
of the temperature range toward both lower and higher values. The harder 
component
of the three is not very well constrained, with an uncertainty on the
upper limit $\sim 100\%$ on the best--fit value. In any case, the range
spanned by the three temperatures seems not very large, from 5 to 15
million degrees, compared to the RS CVn and Algol binaries case.
In our analysis, we also find that the RS model yields lower abundances and
temperatures than the MK model. The differences are not very large, but are
consistently present in the analysis of all detectors.

In conclusion, our data show that: i) HD\,35850, a single F-type star,
is a very active X-ray source; ii) its coronal plasma is characterised by
low metal abundance values; iii) its X-ray emission is dominated by a
coronal plasma with temperatures in the range of 5--15 million degrees.

\begin{acknowledgements}
GT and RP acknowledge partial support from the Italian Space Agency.
FH thanks the Observatory of Merate for hospitality, and acknowledges 
financial support by the Swedish Natural Science Research Council.
TAF acknowledges partial support from NASA grant NAG5-2955 under the
ASCA guest observer program. We would like also to thank the referee
Dr. R. Mewe for his useful comments.
\end{acknowledgements}

\end{document}